\documentstyle[aps,prl,multicol,epsf]{revtex}
\begin{document}

\newcommand{\tbox}[1]{\mbox{\tiny #1}} 
\newcommand{\half}{\mbox{\small $\frac{1}{2}$}} 
\newcommand{\mpg}[2]{\begin{minipage}[t]{#1cm}{#2}\end{minipage}}	
\newcommand{\mpb}[2]{\begin{minipage}[b]{#1cm}{#2}\end{minipage}}
\newcommand{\mpc}[2]{\begin{minipage}[c]{#1cm}{#2}\end{minipage}}

%%%%%%%%%%%%%%%%%%%%%%%%%%%%%%%%%%%%%%%%%%%%%%%%%%%%%%%%%%%%%%%

\title{Quantum dissipation due to the interaction with chaotic 
degrees-of-freedom and the correspondence principle}

\author{Doron Cohen}

\date{October 1998, revised April 1999}

\address{Department of Physics, Harvard University} 

\maketitle

%%%%%%%%%%%%%%%%%%%%%%%%%%%%%%%%%%%%%%%%%%%%%%%%%%%%%%%%%%%%%%

\begin{abstract}
Both in atomic physics and in mesoscopic physics it is 
sometimes interesting to consider the energy time-dependence 
of a parametrically-driven chaotic system. We assume an 
Hamiltonian ${\cal H}(Q,P;x(t))$ where $x(t)=Vt$. 
The velocity $V$ is slow in the classical sense but not 
necessarily in the quantum-mechanical sense. 
The crossover (in time) from ballistic to diffusive 
energy-spreading is studied. The associated irreversible 
growth of the average energy has the meaning of dissipation. 
It is found that a dimensionless velocity $v_{PR}$ determines 
the nature of the dynamics, and controls the route towards 
quantal-classical correspondence (QCC). A perturbative regime 
and a non-perturbative semiclassical regime are distinguished. 
\end{abstract}

%%%%%%%%%%%%%%%%%%%%%%%%%%%%%%%%%%%%%%%%%%%%%%%%%%%%%%%%%%%%%%%%%%%%%%%%
\begin{multicols}{2}

Consider a system that is described by an Hamiltonian
${\cal H}(Q,P;x)$ where $(Q,P)$ are canonical variables and 
$x$ is a parameter. For example, $x$ may represent the position 
of a large object (`piston' or `Brownian particle') 
which is located inside a cavity, and the $(Q,P)$ variables 
may describe the motion of one or few `gas particles'. 
Assume that for $x(t){=}\mbox{\it const}$ the 
motion $(Q(t),P(t))$ is classically {\em chaotic}, 
and characterized by a correlation time $\tau_{\tbox{cl}}$.
Just for simplicity of the following presentation 
one may identify $\tau_{\tbox{cl}}$ with the ergodic time$^{\dag}$.  
In this letter we are interested in the time-dependent 
case where $\dot{x}{=}V$. The velocity V is assumed to be 
slow in the classical sense. It is not necessarily slow 
in the quantum-mechanical (QM) sense. The notion of slowness 
is an important issue that we are going to discussed 
in detail.  
Because $V {\ne} 0$, the energy is not a constant-of-motion.  
We study the crossover (in time) from ballistic to diffusive 
{\em energy-spreading}, and the associated irreversible growth of 
the average energy ${\mathsf E}=\langle{\cal H}\rangle$. 
By definition, this growth has the meaning of {\em dissipation}.  
It is common to {\em define} the dissipation coefficient 
$\mu$ via $\dot{{\mathsf E}}=\mu V^2$, where $\dot{{\mathsf E}}$ is 
the dissipation rate. 
The correspondence between quantal dissipation 
and classical dissipation should not be taken 
as obvious. It is expected that in the 
$\hbar\rightarrow 0$ limit the quantal $\mu$ will 
become similar to the classical $\mu$.  However, 
this is just an expectation.  The actual `proof' 
should be based on proper implementation of 
QM considerations.  This considerations 
should further clarify what does it mean to have 
small $\hbar$, and what happens if $\hbar$ is not 
very small. The clarification of these issues is 
the main purpose of the present letter.

The interest in quantum-dissipation is very old 
\cite{bath,wall,koonin}. 
There are few approaches to the subject. The most 
popular is the {\em effective-bath approach} \cite{bath}. 
When applied to `our' problem (as defined above) it means 
that the chaotic $(Q,P)$ degrees-of-freedom are replace by an 
effective-bath that has the same {\em spectral-properties}. 
This may be either harmonic-bath 
(with infinitely many oscillators) or 
random-matrix-theory (RMT) bath. It turns out 
that quantal-classical correspondence (QCC) is a natural 
consequence of this procedure: The dissipation 
coefficient $\mu$ turns out to be the same classically 
and quantum-mechanically. The effective-bath approach will 
not be adopted in this letter since its applicability 
is a matter of conjecture. We want to have a direct understanding 
of quantum-dissipation.

The understanding of {\em classical} dissipation,  
in the sense of this letter, is mainly based 
on \cite{wall,koonin}. 
Quantum-mechanically much less is known. 
Various {\em perturbative methods} have been used 
\cite{koonin,wilk,berry} in order to obtain an 
expression for the quantum-mechanical $\mu$.  
These methods are (essentially) variations 
of the well known Fermi-Golden-Rule (FGR). 
The {\em simple} FGR expression for $\mu$ does not 
violate the expected correspondence with the classical 
result. However, this has been challenged most clearly 
by Wilkinson and Austin (W\&A) \cite{wilk}. 
They came up with a surprising conclusion 
that we would like to paraphrase as follows: 
A proper FGR picture, supplemented by an innocent-looking RMT 
assumption, leads to a {\em modified} FGR expression;  
In the $\hbar\rightarrow 0$ the modified FGR expression 
disagrees with the classical result. This observation was 
the original motivation for the present study.

The outline of this letter is as follows: 
In the next paragraph we are going to give a terse outline 
of our main observations. Then we start with a brief review 
of the classical picture. Most importantly:  
{\em It should be realized that the analysis of 
dissipation is reduced to the study of energy spreading}. 
This observation is valid classically as well as quantum-mechanically, 
and constitutes the corner-stone in the derivation of the 
universal fluctuation-dissipation (FD) relation.  
The rest of the paragraphs are dedicated to the presentation 
of the QM considerations. We are 
going to establish quantal-classical correspondence (QCC)  
in the limit $\hbar\rightarrow 0$ using semiclassical 
considerations. The detailed discussion of W\&A's  
RMT approach is deferred to a long paper \cite{frc}.

The main object to be discussed in this letter is the 
transition probability kernel $P_t(n|m)$. The variable $m$ denotes 
the initial energy preparation of the system. 
It is a level index in the QM case.  
After time $t$ the parameter $x{=}x(0)$ has a new value 
$x{=}x(t)$.  Therefore it is possible to define a new set 
of (instantaneous) energy-eigenstates that are labeled 
by the index $n$.  Thus, the kernel $P_t(n|m)$, regarded  
as a function of~$n$, describes an evolving energy-distribution. 
One may wonder whether the quantum-mechanical 
$P_t(n|m)$ is similar to the corresponding 
classical object. We shall distinguish between 
{\em detailed} QCC and {\em restricted} QCC. 
The latter term implies that only the second-moment 
of the spreading profile is being considered. 
The crossover from ballistic to diffusive 
energy-spreading happens at $t\sim\tau_{\tbox{cl}}$. 
In order to {\em capture} this crossover within quantum-mechanics, 
a proper theory for the quantal $P_t(n|m)$ should be constructed. 
We shall define a scaled-velocity $v_{\tbox{PR}}$.  
Our first main observation is that $v_{\tbox{PR}} \ll 1$ 
is a necessary condition for the applicability 
of {\em perturbation theory}. In the perturbative regime 
the quantal $P_t(n|m)$ is {\em not} similar to the classical 
$P_t(n|m)$, there is no detailed QCC, but still one can 
establish {\em restricted} QCC.  
If $v_{\tbox{PR}} \gg 1$, then the crossover 
at $t\sim\tau_{\tbox{cl}}$ is out-of-reach for perturbation theory. 
Consequently non-perturbative approach is essential.  
This turns out to be the case in the limit $\hbar\rightarrow 0$. 
Our second main observation is that $v_{\tbox{PR}} \gg 1$ 
is a necessary condition for {\em detailed} QCC. 
The latter is the consequence of {\em semiclassical} considerations.

The starting point for the classical theory of 
dissipation \cite{wall,wilk} is the statistical characterization 
of the fluctuating quantity  
${\cal F}(t) = -(\partial {\cal H} / \partial x)$, 
assuming that $x{=}\mbox{\it const}$.  
Without loss of generality$^{\ddag}$ 
it is further assumed that the  
average force is $\langle {\cal F} \rangle_{\tbox{E}} = 0$.  
The angular brackets stand for a microcanonical average 
over $(Q(0),P(0))$, where $E$ is the energy.  
Note again that we still assume $V{=}0$. 
The temporal correlations of the stochastic-like force are 
$C_{\tbox{E}}(\tau)=\langle {\cal F}(t) {\cal F}(t{+}\tau) \rangle$. 
It is assumed that the classical $C_{\tbox{E}}(\tau)$  
is characterized by a correlation time $\tau_{\tbox{cl}}$. 
The intensity of fluctuations is described by 
the parameter $\nu = \tilde{C}(0)$. 
The power-spectrum of the fluctuations 
$\tilde{C}_{\tbox{E}}(\omega)$ is defined 
via a Fourier transform.

For finite $V$ the energy 
${\cal E}(t) = {\cal H}(Q(t),P(t);x(t))$  
is not a constant-of-motion. 
After time $t$ the energy change is simply 
$({\cal E}(t){-}{\cal E}(0)) = -V \int_0^t {\cal F}(t)dt$.
Let us assume that at $t{=}0$ we have a 
{\em microcanonical distribution} of initial `points'. 
For short times $t \ll \tau_{\tbox{cl}}$ one can prove that 
the evolving phase-space distribution is still 
confined to the initial energy surface. Thus the evolving 
distribution remains equal to the initial microcanonical distribution.   
This is the so-called classical {\em sudden approximation}. 
Squaring $({\cal E}(t){-}{\cal E}(0))$ and 
averaging over initial conditions we find that 
for short times we have a ballistic energy spreading: 
\begin{eqnarray} \label{e1}
\langle ({\cal E}(t)-{\cal E}(0))^2 \rangle 
= C(0) \times (Vt)^2 
\end{eqnarray}
This ballistic behavior is just a manifestation of 
the parametric energy change 
$\delta{\cal E}=(\partial {\cal H} / \partial x) \delta x$. 
For longer times ($t \gg \tau_{\tbox{cl}}$) 
we get a diffusive energy spreading: 
\begin{eqnarray} \label{e2}
\langle ({\cal E}(t) & - & {\cal E}(0))^2 \rangle 
\ \approx \ 2D_{\tbox{E}} t  
\\ \label{e3} 
D_{\tbox{E}} \ & = & \ \half \ \nu \ V^2
\end{eqnarray}
Thus (for $t \gg \tau_{\tbox{cl}}$)  
the evolving phase-space distribution is 
concentrated around the evolving 
energy surface ${\cal H}(Q,P;x(t))=E$.
This is the so-called classical {\em adiabatic approximation}. 
It becomes exact if one takes 
(after substitution of (\ref{e3}) into (\ref{e2}))
the formal limit $V\rightarrow 0$, keeping $Vt$ constant. 
It should be evident that for finite $V$ there 
is eventually a breakdown of the adiabatic approximation. 
The time $t_{\tbox{frc}}$ of this breakdown is estimated 
in the next paragraph.    
The only approximation that was involved in the 
above analysis is that 
$\langle {\cal F}(t) {\cal F}(t{+}\tau) \rangle 
\approx C_{\tbox{E}}(\tau)$.  A strict equality 
applies (by definition) only if $V{=}0$.  Detailed discussion of 
this approximation is quite straightforward \cite{frc}. 
It leads to the {\em classical slowness condition}. 
For the `piston' example one easily concludes that 
the velocity $V$ of the piston should be much smaller 
compared with the velocity of the gas particle(s).

The diffusion across the evolving 
energy surface leads to an associated systematic 
growth of the average energy~${\mathsf E}$. 
This is due to the $E$-dependence of $D_{\tbox{E}}$.  
The rate of energy growth is $\dot{{\mathsf E}}=\mu V^2$. 
The dissipation coefficient is related 
to the intensity of the fluctuations as follows:
\begin{eqnarray} \label{e4}  
\mu \ = \ \frac{1}{2} \frac{1}{g(E)} 
\frac{\partial}{\partial E}
(g(E) \nu)
\end{eqnarray}
Here $g(E)=\partial_{\tbox{E}}\Omega(E)$ 
is the classical density of states and   
$\Omega(E)$ is the phase-space volume which 
is enclosed by the respective energy surface. 
A canonical energy-averaging over the 
above FD relation leads to the familiar form 
$\mu=\nu/(2k_BT)$, where $T$ is the temperature.
The irreversible growth of the average energy 
($\dot{{\mathsf E}}{=}\mu V^2$) implies that the 
fluctuating quantity ${\cal F}(t)$ has a non-zero average.  
Namely $\langle {\cal F} \rangle = -\mu V$. 
In case of the `piston' example the latter is 
commonly named `friction' force. 
The classical adiabatic approximation~(\ref{e2}) 
is valid as long as the systematic growth of the 
average energy ($\dot{{\mathsf E}}t$) is much smaller 
than $\sqrt{2D_{\tbox{E}}t}$. It leads to an  estimate 
for the classical breaktime $t_{\tbox{frc}}=\nu/(\mu V)^2$.

In order to make a smooth transition from the 
classical to the QM formulation 
we define the following kernels:  
\begin{eqnarray} \label{e5}  
P_t(n|m) \ & = & \ \mbox{trace}
( \ \rho_{n,x(t)} \ {\cal U}(t) \ \rho_{m,x(0)} \ ) \\
P(n|m) \ & = & \ \mbox{trace}
( \ \rho_{n,x(t)} \ \rho_{m,x(0)} \ ) 
\end{eqnarray}
In the classical context $\rho_{n,x}(Q,P)$ is defined as 
the microcanonical distribution that is supported 
by the energy-surface ${\cal H}(Q,P;x(t))=E_n$. 
The energy $E_n$ corresponds to the phase-space 
volume $n{=}\Omega(E)$. In the QM context 
$\rho_{n,x}(Q,P)$ is defined as the Wigner-function 
that represents the energy-eigenstate $|n(x)\rangle$. 
The phase-space propagator is denoted 
symbolically by ${\cal U}(t)$. In the classical case 
it simply re-positions points in phase-space. 
In the QM case it has a more complicated structure.  
The trace operation is just $dQdP$ integration.  
It is convenient to measure phase-space volume 
($n{=}\Omega(E)$) in units of $(2\pi\hbar)^d$ 
where $d$ is the number of degrees of freedom.  
This way we can obtain a `classical approximation' for 
the QM kernel, simply by making $n$ and $m$ integer 
variables. If the `classical approximation' is  
similar to the QM kernel, then we say that there is 
detailed QCC$^{\S}$. If only the second-moment is 
similar, then we say that there is restricted QCC. 
The parametric kernel $P(n|m)$ is just the 
projection of the initial energy 
surface\mbox{\small /}eigenstate (labeled by $m$) 
on the new set of energy 
surfaces\mbox{\small /}eigenstates (labeled by $n$). 
For the parametric kernel, only the parametric  
change $\delta x \equiv Vt$ is important. The actual 
time ($t$) to realize this change is not important (by definition). 
This is not true for the actual kernel 
$P_t(n|m)$. The latter is defined as the projection 
of an evolving surface\mbox{\small /}eigenstate, 
where $m$ is taken as the  initial preparation. 
In the QM case we may use more conventional 
notation and write 
$P_t(n|m) = |{\mathbf U}_{nm}(t)|^2$    where 
${\mathbf U}_{nm}(t) = \langle n(x(t))| {\mathbf U}(t) | m(0) \rangle$  
are the matrix elements of the evolution operator. 
Similarly, the parametric kernel $P(n|m)$ is related to 
the transformation matrix 
${\mathbf T}_{nm}(x) = \langle n(x)| m(0) \rangle$.

Let us paraphrase the classical discussion that 
leads to (\ref{e4}). By definition, the departure 
of $P_t(n|m)$ from $P(n|m)$ marks the breakdown 
of the sudden approximation. The spreading 
of $P_t(n|m)$, which is implied by~(\ref{e2}), 
reflects the deviation from the 
(classical) adiabatic approximation.  
The stochastic nature of the spreading on long times 
implies a systematic growth of the average energy. 
The considerations that lead to the 
FD relation~(\ref{e4}) are of quite 
general nature, provided {\em genuine} 
diffusive behavior is established: It is argued \cite{frc} 
that $P_t(n|m)$ can be written as the convolution 
of $N$ kernels $P_{t_1}(n|m)$ such that $t=Nt_1$ 
and $\tau_{\tbox{cl}} \ll t_1 \ll  t_{\tbox{frc}}$. 
Using general argumentation, the same as in the 
derivation of the `central limit theorem', 
one concludes that $P_t(n|m)$ obeys a diffusion 
equation. The diffusion coefficient $D_{\tbox{E}}$ is 
determined by the second-moment of $P_{t_1}(n|m)$, 
and hence it is given by~(\ref{e3}). 
Most importantly, higher-moments 
of $P_{t_1}(n|m)$ become irrelevant.
The validity of the above argumentation is conditioned 
by the requirement of having the separation of time scales 
$\tau_{\tbox{cl}} \ll t_{\tbox{frc}}$.  This is a main ingredient 
in the {\em classical definition of slowness}. 
In the QM case we can try  
to use the same reasoning in order to derive 
a FD relation that corresponds to~(\ref{e4}). 
The crucial step is to establish the diffusive 
behavior~(\ref{e2}) for a limited time scale 
which is required to be much longer than $\tau_{\tbox{cl}}$. 
The quantum-mechanical $D_{\tbox{E}}$ will hopefully
corresponds to~(\ref{e3}). This correspondence is 
not a-priori guaranteed. This is the main issue  
that we are going to address in the rest of this letter.  
First we discuss the conditions for having 
QCC for the parametric kernel $P(n|m)$. 
Then we discuss the departure of $P_t(n|m)$ 
from $P(n|m)$. The conditions for having 
either detailed or at least restricted QCC  
will be specified.

The quantal $\rho_{n,x}(Q,P)$, unlike its classical 
version, has a non-trivial transverse structure. 
For a curved energy surface the transverse profile 
looks like Airy function and it is characterized by a width 
$\Delta_{\tbox{SC}} = 
( \varepsilon_{\tbox{cl}} 
(\hbar/\tau_{\tbox{cl}})^2 )^{1/3}$ 
where $\varepsilon_{\tbox{cl}}$ is a classical 
energy scale. For the `piston' example  
$\varepsilon_{\tbox{cl}}{=}E$ is the kinetic 
energy of the gas particle. 
Given a parametric change $\delta x \equiv Vt$ we can define 
a classical energy scale 
$\delta E_{\tbox{cl}} \propto \delta x$ 
via~(\ref{e1}).  This parametric 
energy scale characterizes the transverse 
distance between the intersecting energy-surfaces 
that support $|m(x)\rangle$ and $|n(x{+}\delta x)\rangle$. 
Considering $P(n|m)$, it should be legitimate to 
neglect the transverse profile of Wigner function 
provided $\delta E_{\tbox{cl}} \gg \Delta_{\tbox{SC}}$. 
This condition can be cast into the 
form $\delta x \gg \delta x_{\tbox{SC}}$ where 
$\delta x_{\tbox{SC}}  =   
\Delta_{\tbox{SC}} /
(\sqrt{\nu^{\tbox{cl}}/\tau_{\tbox{cl}}})$. 
If $\delta x \ll \delta x_{\tbox{SC}}$ we cannot 
argue that there is detailed QCC. On the 
other hand we can not rule out such QCC. 
We shall come back to this issue shortly.

If $\delta x$ is sufficiently small it should be 
possible to get an approximation for ${\mathbf T}_{nm}(x)$, 
and hence for $P(n|m)$, via perturbation theory. 
It turns out that in the perturbative regime $P(n|m)$ 
is characterized by a core-tail structure that does not 
correspond to the classical $P(n|m)$. 
Detailed definition and discussion 
of the core-tail structure \cite{frc} are not required 
for the following considerations. The only important 
(non-trivial) observation is that in-spite of the 
lack of detailed QCC there is still restricted QCC. 
Namely, the second moment is still 
given by  $\delta E_{\tbox{cl}}$. An estimate 
for the breakdown of perturbation theory can be 
easily obtained using heuristic considerations as follows:   
The range of first-order transitions is determined 
by the bandwidth $\Delta_b=\hbar/\tau_{\tbox{cl}}$ 
of the matrix $({\partial {\cal H}}/{\partial x})_{nm}$. 
See \cite{berry}. In the perturbative regime $P(n|m)$ is 
non-vanishing only if $|E_n-E_m|<\Delta_b$.  
Perturbation theory is inapplicable unless the 
spreading is on an energy-scale  
$\delta E_{\tbox{cl}} \ll \Delta_b$. 
This condition can be cast into the 
form $\delta x \ll \delta x_{\tbox{prt}}$ where 
\begin{eqnarray} \label{e7}
\delta x_{\tbox{prt}} \  =  \  
\hbar \ / \ 
\sqrt{\nu^{\tbox{cl}} \tau_{\tbox{cl}}} 
\end{eqnarray}
We have $\delta x_{\tbox{prt}} \propto \hbar$ and 
$\delta x_{\tbox{SC}} \propto \hbar^{2/3}$. 
Therefore, typically, the two parametric scales are 
well separated: We do not have a theory for the 
intermediate parametric regime 
$\delta x_{\tbox{prt}} \ll \delta x \ll \delta x_{\tbox{SC}}$. 
The only thing that we can say with confidence is 
that the core-tail structure of $P(n|m)$ is 
washed away once $\delta x > \delta x_{\tbox{prt}}$. 
We do not know whether there is an additional 
crossover once we go paste $\delta x_{\tbox{SC}}$. 
Therefore it is more meaningful to state that 
$\delta x > \delta x_{\tbox{prt}}$ is a {\em necessary}  
condition for detailed QCC, rather than specifying 
the {\em sufficient} condition $\delta x > \delta x_{\tbox{SC}}$.

We turn now to discuss the actual transition 
probability kernel $P_t(n|m)$. 
Recall that our objective is to {\em capture} the crossover 
at $t\sim\tau_{\tbox{cl}}$.  Therefore it is essential 
to distinguish between two possibilities: 
If $V\tau_{\tbox{cl}} \ll \delta x_{\tbox{prt}}$ it means that 
the crossover happens in a regime where 
perturbation theory is still valid.  On the other 
hand  if  $V\tau_{\tbox{cl}} \gg \delta x_{\tbox{prt}}$ it means 
that the crossover is out-of-reach for perturbation 
theory. It is then essential to use non-perturbative 
considerations. The sufficient condition for the 
applicability of semiclassical theory is 
$V\tau_{\tbox{cl}} \gg \delta x_{\tbox{SC}}$. 
Thus we come to the conclusion that the following 
generic dimensionless parameter controls QCC: 
\begin{eqnarray} \label{e8}  
v_{\tbox{PR}} \ = \ \mbox{scaled velocity} \ = \ 
\sqrt{D_{\tbox{E}}\tau_{\tbox{cl}}} / \Delta_b
\end{eqnarray}
If $v_{\tbox{PR}}\ll 1$ then it is feasible to 
extend perturbation theory beyond $\tau_{\tbox{cl}}$. 
This issue is discussed in the next paragraphs.   
In the limit $\hbar\rightarrow 0$ we have 
$v_{\tbox{PR}}\gg 1$ and perturbation theory is inapplicable. 
However, if $v_{\tbox{PR}}$ is sufficiently large 
then semiclassical consideration can be used 
in order to argue that the classical result is valid 
also in the QM domain.  
Before we go on, we should mention an additional  
restriction on QCC, that pertains to $P_t(n|m)$. 
The evolving (classical) distribution 
${\cal U}(t)\rho_{m,x(0)}$ becomes more and more 
convoluted as a function of time. 
This is because of the mixing behavior that  
characterizes chaotic dynamics. For $t>t_{\tbox{scl}}$ 
the intersections with a given instantaneous energy 
surface $n$ become very dense, 
and no-longer can be resolved by quantum-mechanics. 
The semiclassical breaktime $t_{\tbox{scl}}$ is related 
to the failure of the stationary phase approximation \cite{heller}.
In order to establish the crossover from 
ballistic to diffusive energy spreading 
using a semiclassical theory we should satisfy 
the condition $\tau _{\tbox{cl}} < t_{\tbox{scl}}$. 
This velocity-independent condition turns out to be  
not very restrictive \cite{heller}, and we can safely 
assume that it is typically satisfied.

We are now left with the question whether 
restricted QCC is maintained in the 
perturbative regime for $t>\tau_{\tbox{cl}}$. 
Indeed, if $v_{\tbox{PR}}\ll 1$, perturbation 
theory can be used in order to get an approximation 
for ${\mathbf U}_{nm}(t)$ and hence for the 
respective kernel $P_t(n|m)$. 
Then it is possible to calculate the second-moment
of the energy-spreading and to obtain an expression  
that looks like~(\ref{e2}). Expression~(\ref{e3}) 
for $D_{\tbox{E}}$ applies, provided $\nu$ is replaced  
by an effective noise intensity: 
\begin{eqnarray} \label{e9}
\nu^{\tbox{eff}} \ = \ 
\int_{-\infty}^{\infty}\frac{d\omega}{2\pi} 
\tilde{C}_{\tbox{E}}(\omega) \tilde{F}(\omega)
\end{eqnarray}
The detailed derivation of this result 
will be presented elsewhere \cite{frc}.  Formally 
it looks exactly the same as either the simple 
FGR result \cite{koonin} or W\&A's result \cite{wilk}. 
But this formal similarity is quite misleading. The 
difficult issue is how $\tilde{F}(\omega)$ 
is defined.  This function describes the 
{\em effective} power spectrum of the driving force. 
It is the Fourier transform of a correlation 
function $F(\tau)$ with the convention $F(0){=}1$.
The latter is characterized by a correlation 
time $\tau_c$. 
The demonstration that the intrinsic $\tau_c$ 
is much larger than $\tau_{\tbox{cl}}$ is 
the main challenge of the theory \cite{frc}. 
The implied {\em restricted} QCC is explained below.

We turn to discuss the physical consequences 
of~(\ref{e9}). For this purpose we should first explain 
how the QM fluctuation-spectrum look like. 
The fluctuating quantity ${\cal F}(t)$ should 
be handled as an operator. The quantal $C_{\tbox{E}}(\tau)$ is 
similar to the classical $C_{\tbox{E}}(\tau)$ provided 
$\tau\ll t_{\tbox{H}}$. See \cite{berry}. 
Here $t_{\tbox{H}}=\hbar/\Delta$ is the Heisenberg 
time and $\Delta$ is the mean level spacing. 
The associated $\tilde{C}_{\tbox{E}}(\omega)$ 
reflects the discrete nature of the energy levels,  
but its envelope is classical-like.  
The {\em conditions for restricted} QCC are obtained 
by inspection of~(\ref{e9}):  
The function $C_{\tbox{E}}(\tau)$ is characterized by two 
distinguished time scales, which are $\tau_{\tbox{cl}}$ 
and $t_{\tbox{H}}$. The function $F(\tau)$ is characterized 
by a single time scale $\tau_c$.     
Accordingly we have the following possibilities:  
If $\tau_c \ll \tau_{\tbox{cl}}$ then the transitions 
are {\em band-limited} and we get 
$\nu^{\tbox{eff}} \approx 
(\tau_c/\tau_{\tbox{cl}}) \times \nu^{\tbox{cl}}$. 
If $\tau_c \gg \tau_{\tbox{cl}}$ then the transitions 
are {\em resonance-limited} and we get 
$\nu^{\tbox{eff}}\approx\nu^{\tbox{cl}}$. 
In the latter case $\tilde{F}(\omega)$ is essentially 
like a delta function. However $\tilde{F}(\omega)$  
should not be too narrow. Namely, if $\tau_c \gg t_{\tbox{H}}$ 
then the effective noise intensity becomes vanishingly small. 
This is because individual levels are resolved. 
The condition $\tau_c \gg t_{\tbox{H}}$ is satisfied 
only for extremely-slow velocities. This is the so-called  
QM-adiabatic regime. There Landau-Zener transitions are 
the ultimate mechanism for energy-spreading and dissipation 
\cite{wilk}, and QCC is not a-priori guaranteed. 

%%%%%%%%%%%%%%%%%%%%%%%%%%%%%%%%%%%%%%%%%%%%%%%%%%%%%%%%%%%%%%%%%%%%% 

I thank Eric Heller and Shmuel Fishman for stimulating discussions.

%%%%%%%%%%%%%%%%%%%%%%%%%%%%%%%%%%%%%%%%%%%%%%%%%%%%%%%%%%%%%%%%%%%%% 

%%%%%%%%%%%%%%%%%%%%%%%%%%%%%%%%%%%%%%%%%%%%%%%%%%%%%%%%%%%%%%%%%%%%
\end{multicols}
%%%%%%%%%%%%%%%%%%%%%%%%%%%%%%%%%%%%%%%%%%%%%%%%%%%%%%%%%%%%%%%%%%%%

\begin{thebibliography}{99}

\bibitem[\dag]{ft1} By definition the correlation 
time $\tau_{\tbox{cl}}$ pertains to ${\cal F}(t)$ 
which is defined later. In specific examples it may 
be smaller then the ergodic time. This is the case with 
the `piston' example. Assuming that successive collisions 
with its faces are uncorrelated, $\tau_{\tbox{cl}}$ is just 
the duration of a collision. 
(For hard walls $\tau_{\tbox{cl}} \sim 0$).  
On the other hand the ergodic time is 
related to the ballistic time.   

\bibitem[\ddag]{ft2} 
Without loss of generality we assume $\langle {\cal F} \rangle =0$. 
This is equivalent to having phase-space volume $\Omega(E,x)$ which 
is independent of $x$. 
(Such is the case for a gas particle which is affected by 
collisions with a small rigid body that is translated 
inside a large cavity). 
A non-zero average component of $(\partial {\cal H}/\partial x)$ 
corresponds to a conservative force, and should be subtracted from 
the definition of ${\cal F}$. 

\bibitem[\S]{ft3} 
`Detailed QCC' does not mean 
complete similarity.  The classical kernel is 
typically characterized by various non-Gaussian 
features, such as sharp cutoffs, delta-singularities 
and cusps. These features are expected to be 
smeared in the QM case. 
The discussion of the latter issue is beyond 
the scope of the present letter.

\bibitem{bath} 
R. Beck and D.H.E. Gross,Phys. Lett. {\bf 47}, 143 (1973). 
\ R. Zwanzig, J. Stat. Phys. {\bf 9}, 215 (1973).
\ D.H.E. Gross, Nuclear Physics {\bf A240}, 472 (1975)
\ A.O. Caldeira and A.J. Leggett,
Physica {\bf 121 A}, 587 (1983).
\ A. Bulgac, G.D. Dang and D. Kusnezov, 
Phys. Rev.E {\bf 58}, 196 (1998). 
\ D. Cohen, Phys. Rev. Lett. {\bf 78}, 2878 (1997).  
\ D. Cohen, J. Phys. A {\bf 31}, 8199-8220 (1998).

\bibitem{wall}
\ J. Blocki, Y. Boneh, J.R. Nix, 
J. Randrup, M. Robel, A.J. Sierk and W.J. Swiatecki, 
Ann. Phys. {\bf 113}, 330 (1978). 
\ E. Ott, Phys. Rev. Lett.{\bf 42}, 1628 (1979). 
\ C. Jarzynski, Phys. Rev. Lett. {\bf 74}, 2937 (1995).

\bibitem{koonin} 
S.E. Koonin, R.L. Hatch and J. Randrup, 
Nuc. Phys. A {\bf 283}, 87 (1977). 
\ S.E. Koonin and J. Randrup, 
Nuc. Phys. A {\bf 289}, 475 (1977). 
 
\bibitem{wilk}
M. Wilkinson, J. Phys. A {\bf 21}, 4021 (1988).
\ M. Wilkinson and E.J. Austin, J. Phys. A {\bf 28}, 2277 (1995). 

\bibitem{berry}
M.V. Berry and J.M. Robbins, 
Proc. R. Soc. Lond. A {\bf 442}, 659 (1993).

\bibitem{frc} D. Cohen (preprint). 

\bibitem{heller}
M.A. Sepulveda, S. Tomsovic and E.J. Heller,
Phys. Rev. Lett. {\bf 69}, 402 (1992). 
B.V. Chirikov, chao-dyn/9705003, chao-dyn/9705004. 

\end{thebibliography}
\end{document}